\documentstyle[prl,twocolumn,aps]{revtex}
\input{epsf}

\newcommand{\bq}{\begin{equation}}
\newcommand{\ee}{\end{equation}}
\newcommand{\fr}[2]{\frac{#1}{#2}}
\newcommand{\eps}{\varepsilon}

\begin{document}
\draft

\title{ ``Lifshitz tails'' and extended states in an Imaginary
random potential}

\author{P.G.Silvestrov}

\address{Budker Institute of Nuclear Physics, 630090
Novosibirsk, Russia}

\maketitle
\date{today}

\begin{abstract}

Non-Hermitean operators may appear 
during the calculation of a partition function in various models
of statistical mechanics.
The tail eigen-states, having anomalously small real
part of energy $Re(\eps)$, became naturally important in this
case. We consider the distribution of such states and the form
of eigenfunctions for the particle propagating in an imaginary
random potential (the model motivated by the statistics of
polymer chains). Unlike it is 
in the Hermitean quantum mechanics, our tail states 
are sufficiently extended. Such state appear if the values of
random potential turns out to be anomalously close inside the
large area. Results of numerical simulations in the case
of strong coupling confirm the analytic estimates.

\end{abstract}


The phenomena of Anderson localization~\cite{Anderson} of a
quantum particle in the random potential attracted a permanent
interest during last 40 years. On the other hand, a considerable
attention have been paid last years to the investigation of
features of eigenfunctions of non-Hermitean Hamiltonians with
disorder\cite{2,3,4,5,Hatano,fse,YaPRBPRL,HatNel,simons1,simons2}.
A popular example of this kind is the Hatano-Nelson model
\cite{Hatano}, where in the presence of the imaginary
vector-potential the transition between real and complex
spectrum takes place \cite{Hatano,fse}. This transition is
usually referred as the mobility edge in 1d problem (which is
already surprising\cite{note0}). In this paper we consider
another example, where the localized(delocalized) eigenfunctions
of non-Hermitean Hamiltonian behave in the "inverted", compared
to the Hermitean case, way.

Two recent papers \cite{simons1,simons2} deals with the
non-Hermitean quantum mechanical Hamiltonian
with an imaginary random potential
\bq\label{hamS}
H=\fr{p^2}{2m}+iV(r) \ .
\ee
This study was motivated by
the observation that
the Euclidean evolution operator $\langle r|
\exp [-tH]|0\rangle$ with $H$~(\ref{hamS}) after averaging over
$\delta$-correlated disordered potential $V$ coincides with the
probability distribution $Z(r,t)$ for the Edwards self-repulsing
polymer \cite{Edwards}.

Many applications of non-Hermitean operators (like both our
examples) come from the statistical physics and naturally deal
with the imaginary time(Euclidean) evolution. In its turn, for
the Euclidean 
evolution operator the ground state contribution (contribution
from the states with lowest $Re\eps$) is naturally
enhanced~\cite{note}. Therefore in this paper we are going to
consider the tail of the eigenfunctions
of the Hamiltonian~(\ref{hamS}), having smallest real part of
energy $Re\eps$. In the Hermitean quantum
mechanics eigen-states with anomalously small $\eps$ (so called
``Lifshitz tails''~\cite{Lifshitz}) originates from the rare
localized fluctuations of the disorder $V$ and are themselves
well localized. For the Hamiltonian eq.~(\ref{hamS}) first of
all the energy is bounded from below $Re\eps>0$~\cite{not}. The
most surprising is the fact that for smaller and smaller
$Re\eps$ in eq.~(\ref{hamS}) (closer to $Re\eps=0$) one finds
(exponentially rare) the more and more extended states.
Investigation of this ``delocalization at the tail'' in
imaginary disordered potential will be the main goal of this
paper\cite{endend}.

In all cases the tail states appear only due to some special
rare fluctuations of the random potential. The
standard procedure is to derive some kind of saddle-point
equation in order to find the most probable fluctuation giving
rise to the eigenmode with given (small) energy. Both the
potential and the wave function in this case are the
instanton-like solutions of some nonlinear differential
equation~\cite{Lifshitz,simons2}. This straight scenario breaks
up for the model eq.~(\ref{hamS}). The necessary fluctuation in
our case is simply $V\approx const$ inside some sufficiently
large connected area. The small probability to have no
considerable fluctuations on top of this $V=const$ leads to the
exponential suppression of density of states at the tail. This
absence of the well defined saddle-point equation describing
the rare tail states may not be specific only for our
Hamiltonian with random imaginary potential. It will be
extremely interesting to find another physically motivated
models, where the Instanton is not defined via the solution of
any classical equation of motion.

In this paper we will investigate in detail only the case of
strong disorder. To this end it is natural to consider the
discretized version of the Hamiltonian~(\ref{hamS})
\bq\label{ham}
(H\psi)_n=-t(\psi_{n+1} + \psi_{n-1}) +iV_n\psi_n =\eps \psi_n \
\ .
\ee
This is for 1-dimension. Generalization for the arbitrary
dimension $d$ is straightforward. Let random $V$ be uniformly
distributed over $-w/2<V_n<w/2$ with
$w\gg t$. Taking into account Gaussian distributed disorder is
important if one would like to consider the mapping onto some
kind of polymer problem. In our case however, Gaussian potential
will only introduce unnecessary complexifications like
nonuniform distribution of eigenvalues along the imaginary axis.
In the strong coupling limit states in the bulk of spectrum are
evidently localized, making it easy to distinguish the extended
states at the border.  For the continuous model (\ref{hamS})
disorder also should be always considered as strong for the
small enough $Re\eps$. Therefore although the features of
solutions of the eq.~(\ref{ham}) will be very peculiar in the
bulk, we still may expect some kind of universal behavior at the
tail.

In the leading approximation in $t/w$ 
one has
\bq\label{e0}
\psi
_n\approx \delta_{n,m} \ , \ \eps
\approx iV_{m} \ .
\ee
The wave function is localized at the site $m$ and the energy is
pure imaginary. Corrections to the energy appear only in the second
order in $t$ (again the generalization for arbitrary $d$ is trivial) 
\bq\label{e1}
\eps
=iV_{m} -\fr{it^2}{V_{m}-V_{m+1}}
-\fr{it^2}{V_{m}-V_{m-1}} +... \ .
\ee 
There is also no real part of $\eps$. Moreover since the
perturbation ($t$) has no diagonal matrix elements the
corrections to energy may be only of the even order in $t$ and
consequently odd in $iV$. Therefore, in any order of
perturbation theory the energy is pure imaginary. The nonzero
$Re\eps$ appears if the values of potential at two neighboring
sites occasionally turns out to be very close $V_1-V_2\sim t$.
The two values of energy are now found by simple diagonalization
of $2\times 2$ matrix
\bq\label{eex2}
\eps=i\fr{V_1+V_2}{2} \pm \sqrt{t^2-\fr{(V_1-V_2)^2}{4}} \ \ .
\ee
For $|V_1-V_2|>2t$ the energy is complex, but the
perturbative series in $t/(V_1-V_2)$ has finite radius of
convergence.  For our model (\ref{ham}) the $4t
d/w$ part of states are beyond this convergence region and
acquire the nonzero $Re\eps$. The nonperturbative origin of
$Re\eps$ is clearly an artifact of the strong coupling
limit\cite{note}.  However, it will teach us how the extended
states may appear in this model. On fig.~1 we have plotted the
distribution of real part of energy for the $1d$ and $2d$
versions of the Hamiltonian (\ref{ham}) found numerically for
different values $w/t$.

In agreement with our prediction the huge $\delta$-peak appears
at $Re\eps=0$. Due to the vanishing of the derivative
$dRe\eps/d(V_1-V_2)$ at $V_1=V_2$ the density of states
described by the eq.~(\ref{eex2}) is singular at $Re\eps=\pm t$
(rainbow singularity) 
\bq\label{rainbow}
\eps\equiv x+iy \ , \ 
P(x)=\fr{2d|x|}{w\sqrt{x^2-t^2}} \ .
\ee
Here we use $x$ and $y$ for real and imaginary part of energy.
The two peaks at $x=\pm t$ are clearly seen at the figure. (The
data for smallest values of $w/t$ corresponds more to the case
of intermediate disorder. Still some sharp peaks at $x$ close to
$\pm t$ are seen in this case.)

In the case of only two close states
~(\ref{eex2}) one has $-t<x<t$. However, if where will be no
disorder at all $x$ will vary within
$-2dt<x<2dt$. These two examples show, how to construct the
fluctuations of $V$ which may lead to the smallest values of
$x$. One should simply arrange a big cluster with anomalously
close values of $V$ at many neighboring sites of the lattice.
In $1d$ the cluster means a chain of
$q$ consecutive sites with close $V_n$, $V_i\approx
V_{i+1}\approx ...\approx V_{i+q-1}\approx \overline{V}$. The
Schr\"{o}dinger equation (\ref{ham}) on a finite segment has $q$
solutions of the form ($1\le k\le q$)
\begin{eqnarray}\label{psiq}
\psi_k^q(n)=\sqrt{\fr{2}{q}}\sin\left( \fr{\pi kn}{q+1}\right) \
&;& 
0<n<q+1 \ , \nonumber\\
\eps_k= -2t\cos\left( \fr{\pi
k}{q+1}\right)&+&\fr{i}{q}\sum_{m=1}^q 
V_m+ {\cal O}(V^2/t)
\ \ .
\end{eqnarray}
Here $|V_n-\overline{V}| \ll t$ for all $1\le n\le q$. The tail
states correspond to $k=1$. For $q=2$, $k=1$ and $|V_1-V_2|\ll
t$ the eq.~(\ref{psiq}) reduces to the eq.~(\ref{eex2}). The
solutions with $k>1$ are also extended, but they have large
$Re\eps$. Therefore, $\psi_k$ with $k>1$ at given $\eps$ 
are statistically insufficient (except for the highest
state with $k=q$).
Taking into account the mixing with sites $0$ and $q+1$ may be
done perturbatively. However, now the correction to energy due
to the $V_{0(q+1)}$ is
pure imaginary in any order of perturbation theory.
The only corrections to the real part of energy described by the
eq.~(\ref{psiq}) comes from $V_n$ with $1\le n\le q$.

\vspace{-.8cm}
\begin{figure}[t]
\epsfxsize=8.8cm
\epsffile{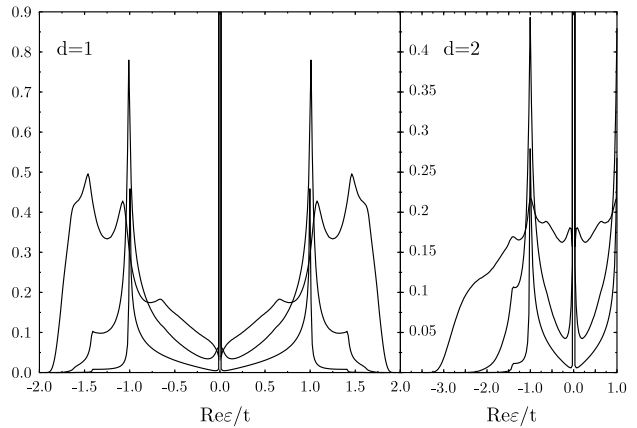}
\vglue 0.2cm
\medskip
\caption{The distribution function $P(Re\eps)$
for one and two spatial dimensions. From top to bottom
$w/t=4,16,64$ for $d=1$ and
$w/t=8,32,128$ for $d=2$. Due to the symmetry
$P(x)=P(-x)$ only a part of positive $x$ region is shown for
$d=2$. $\delta$-function at $x=0$, $(x\pm t)^{-1/2}$ singularity
and the jump at $x=\pm\sqrt{2}t$ are seen at large $w/t$
for both $d=1,2$.}
\end{figure}

The most interesting in the eq.~(\ref{psiq}) is that
while looking for the state with small $ Re\eps$ we have got in
addition the smooth eigenfunction extended over the large ($q\gg
1$) number of lattice sites. (We prefer not to call our states
delocalized since the localization length $\xi$ is usually
defined via the exponent governing the decay of the tail of wave
function and in our case $\xi\ll 1$.) 

On the fig.~2 we have shown the example of a rare extended wave
function for the $1d$ model with $w/t=10$. We have found only
one such "delocalized" eigenmode among $\sim 2\times 10^{9}$
($6\times 10^7$ realizations of the random potential on the
sample having length $l=30$ lattice sites). The size of the
cluster is $q=11$ and $Re\eps=-1.886$. Another quantity, which
may characterize the number of components of the wave function
is the participation ratio $p=(\sum
|\psi(i)|^2)^2/\sum|\psi(i)|^4$. For the function shown on the 
fig.~2 $p=7.62$, which is consistent with $p=2q/3$, following
from the eq.~(\ref{psiq}). 

\vspace{-.8cm}
\begin{figure}[t]
\epsfxsize=8.8cm
\epsffile{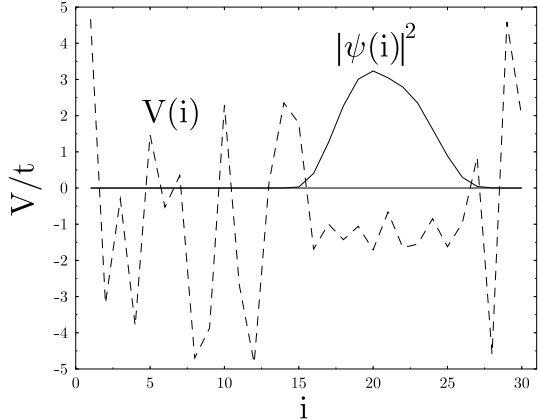}
\vglue 0.2cm
\medskip
\caption{The example of rare fluctuation of the potential $V(i)$
(dashed) for $w/t=10$, giving rise to the extended state.
Solid line is the $|\psi(i)|^2$, arbitrary units. 
The probability to find such fluctuation is $P\sim 10^{-9}$.}
\end{figure}

We devote the rest of the paper to the estimation of the
probability to find the extended tail states. 

For any $q$ the real part of energy described by the
eq.~(\ref{psiq}) is bounded from below. In particular for
large~$q$  
\bq\label{bound}
\min Re\eps=\eps_0^{(q)}\approx -2t+t\pi^2/q^2 \ \ .
\ee
In the more
realistic case the real part of energy is given by the second
order formula 
\bq\label{real}
\eps(V)^{(q)}=\eps_0^{(q)}+\sum_{n=2}^q
\fr{\langle 1|\hat{V}|n\rangle\langle n|\hat{V}|1\rangle}
{\eps_n-\eps_1} \ .
\ee 
Here the summation goes over the states on the cluster (chain of
the length $q$) described by the eq.~(\ref{psiq}). $\hat{V}$ is
the real potential eq.~(\ref{ham}). The probability to find such
cluster in the leading order may be found by simple counting of
the number of
independent variables $V_n$ entering the eq.~(\ref{real}). This
gives 
\bq\label{Pq}
P^{(q)}\sim \int\delta(\eps(V)^{(q)}-x)
\prod_{i=1}^q\fr{dV_i}{w}\sim
\left[\fr{t(x-\eps_0^{(q)})}{w^2} \right]^{\textstyle
\fr{q}{2}} . 
\ee
Here $q\gg 1$. More precisely $P^{(q)}\sim (x
-\eps_0^{(q)})^{(q-3)/2}$, which gives the rainbow singularity 
(\ref{rainbow}) at $x=-t$ and a finite jump of $P(x)$ at
$x=-\sqrt{2}t$. 
For given $x$ the
$P^{(q)}(x)$~(\ref{Pq}) reaches maximum at $q$ equal to
\bq
q=\pi\sqrt{\fr{t}{x-\eps_0}} \ , \ 
x-\eps_0^{(q)}=\fr{2(x-\eps_0)}{\ln\left(
{w^2}/{t(\eps-\eps_0)}\right)} \ ,
\ee
where $\eps_0=\eps_0^{(q=\infty)}=-2t$. The resulting
probability to find the eigenstate, 
given by this largest $P^{(q)}$, reads
\bq\label{P1}
P(x)\sim
\left(\fr{const \times w^2}{t(x-\eps_0)} 
\right)^{\textstyle 
-\fr{\pi}{2}
\sqrt{\fr{t}{x-\eps_0}}
} .
\ee
Here we keep only the leading contributions and the argument of
the exponent in the r.h.s. is found with only $\sim 1/\ln(w/t)$
accuracy. In $2d$ in order to get the anomalously small $Re\eps$
one should have occasionally close values of $V_i$ within the
large circle (radius $r=\sqrt{q/\pi}$). The analog of the
eq.~(\ref{bound}) now became 
\bq\label{e2d}
\eps_0^{(q)}=-4t+\pi j_0^2 t/q \ ,
\ee
where $j_0\approx 2.4048$ is the first node of the Bessel
function $J_0(x)$. Since we are in the strong coupling regime,
eq.~(\ref{Pq}) is valid for any number of spatial dimensions.
However, due to the another $q$-dependence of $\eps_0^{(q)}$
(\ref{e2d}) the probability now has the form ($\eps_0=-4t$)
\bq\label{P2}
P(x)\sim
\left(\fr{const \times w^2}{t(x-\eps_0)} 
\right)^{\textstyle 
-\fr{\pi j_0^2}{2}
\fr{t}{x-\eps_0}} .
\ee
In general for arbitrary $d$ we expect ($\eps_0=-2dt$)
\bq
\ln(P(x))\sim - \ln\left(\fr{w^2}{t(x-\eps_0)} 
\right)\times
\left( \fr{t}{x-\eps_0}\right)^{d/2} \ .
\ee
On the figs.~3 and 4 we have plotted the $\ln(P(x))$ for
$d=1,2$, found numerically for few values of $w$, as a function
of scaling variable $\chi=(t/(x-\eps_0))^{d/2}$. The agreement
between the experiment and the theoretical formulas~(\ref{P1})
and (\ref{P2}) (found with only $1/\ln$ accuracy!) for large
$w/t$ is rather appreciable.

To conclude,
we have considered the extended states in the strong coupling
limit of the non-Hermitean Hamiltonian with random imaginary
potential. Unlike it is in the case of Hermitean quantum
mechanics, only the extended states survive close to the
border of spectrum (may have anomalously small (large)
$Re(\eps)$). These extended states appear due to the rare
fluctuations of the random potential, where the values of $V$
inside some sufficiently large area turns out
to be close. To the best of our knowledge, this is the first
consideration of the instanton-like effect, where the most
important large fluctuation of the field ($V$) is not found by
the solution of any equation of motion. In the strong coupling
limit the real part of energy appears only non-perturbatively.
This leads to a special form of the density of states with
various sharp features. This singular behavior is clearly seen
at the histogram representing the results of 
numerical simulations. Simple analytical expressions were found
for the density of states at the tail (and correspondingly for
the probability to find the extended states). Even though the
formulas were found only with the logarithmic accuracy, the
theory agrees well with the results of numerical
experiment. 
Extension of our approach for the case of weak disorder requires
a further investigation. It will be also interesting to find
another examples of rare "physical" events described by
non-classical instantons.

This work was supported by RFBR, grant 98-02-17905.

\vspace{-.8cm}
\begin{figure}[t]
\epsfxsize=8.8cm
\epsffile{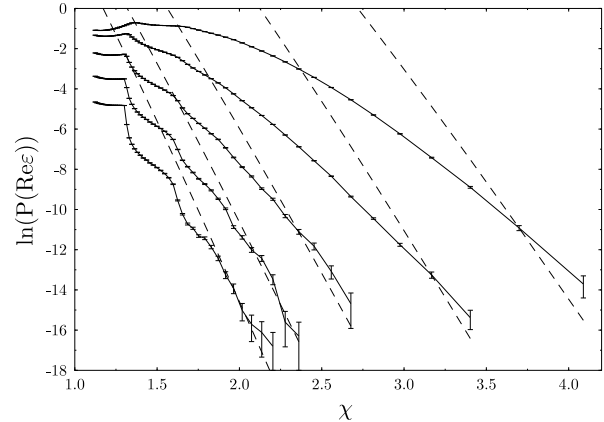}
\vglue 0.2cm
\medskip
\caption{The logarithm of the density of states $\ln
(P(Re\eps))$ as a function of scaling variable
$\chi=[t/(2t+Re\eps)]^{-1/2}$ for the $1d$ Hamiltonian (2).
From left to right $w/t=64,32,16,8,4$. Dashed lines are
the theoretical prediction of eq.~(12) with $const=1$ and
suitable overall normalization.}
\end{figure}

\vspace{-.8cm}
\begin{figure}[t]
\epsfxsize=8.8cm
\epsffile{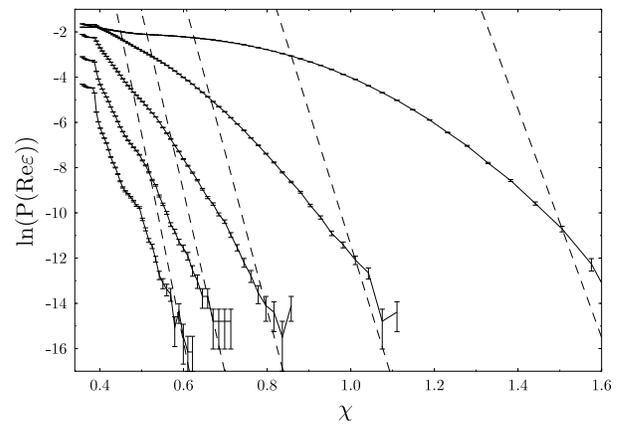}
\vglue 0.2cm
\medskip
\caption{The same as on the fig.~3, but for $2d$. Scaling
variable is $\chi=t/(4t+Re\eps)$, from left to right
$w/t=128,64,32,16,8$. The theory is given by the eq.~(14).} 
\end{figure}


\begin{thebibliography}{99}
\bibitem{note0} The more detailed investigation showed that this
threshold separates the region of localized wave functions from
a very special phase with the solutions demonstrating features
of both localized and delocalized modes~\cite{YaPRBPRL,HatNel}.
\bibitem{note} It is however questionable: does the only
ground state is important? It is so
for the Hatano-Nelson model, where due to this fact all the
reach threshold behavior may become physically relevant only for
ensemble of many repulsing vortices~\cite{YaPRBPRL}. For the
eqs.~(\ref{hamS},\ref{ham}) both the "ground state" energy and
wave function are sufficiently complex. This complexity may lead
to ``exponentially strong'' cancellations while averaging over
$V(r)$.
\bibitem{not} Our tail states (\ref{psiq}) have $Re\eps>0$ by
construction and numerical simulations show the absence of
$Re\eps<0$ for sufficient statistics.
\bibitem{endend} The existence of this states was known to
the authors of ref.~\cite{simons2}, but neither the form of the
wavefunctions, nor the probability to find such rare
fluctuations were investigated. In ref.~\cite{simons2} another
kind of tail states, having anomalously large(small) $Im\eps$
were considered.
\bibitem{note} The finite fraction of states with $Re\eps\equiv
0$ is specific for discretizations of the eq.~(\ref{hamS})
preserving the reflection symmetry of spectrum around the band
center~\cite{simons2}. We performed the numerical simulations
for the discrete Hamiltonian containing both nearest and
next-to-nearest hopping, which has no such symmetry. In this
case the spectrum is completely different from our
fig.~1, but the density of tail states still is described by
the eq.~(\ref{P1}).
\end{thebibliography}
\end{document}